\documentclass[12pt]{article}

\usepackage[english]{babel}
\usepackage{amssymb,latexsym,amsmath}

\begin{document}

\title
{Octonionic instantons in eight dimensions}
\author
{E.K. Loginov\footnote{{\it E-mail address:} ek.loginov@mail.ru}\\
\it Ivanovo State University, Ivanovo, 153025, Russia}
\date{}
\maketitle

\begin{abstract}
In this letter, we study the instanton moduli space of the eight-dimensional solutions of the self-duality equation $F\wedge F= \ast F\wedge F$. Using the known ADHM-construction of such instantons, we compute the dimension of the space of solutions to this construction. We find a general $N$-instanton solution of the 't Hooft type which depends on $8N+1$ free parameters and investigate the cases $N=2$ and $N=3$ in detail.
\end{abstract}

\section{Introduction}

The discovery of regular solutions to the Yang-Mills field equations in the four-dimensional Euclidean space, which correspond to the absolute minimum of the action, has led to an intensive study of such theory and the search multidimensional generalizations of the self-duality equations. In Refs.~\cite{corr83,ward84}, such equations were found and classified. These were first-order equations that satisfy the Yang-Mills field equations as a consequence of the Bianchi identity. Later, solutions to these equations were found and then used to construct classical solitonic solutions of the low energy effective theory of the heterotic string~(see Refs.~\cite{fair84,fubi85,corr85,ivan92,harv91,khur93,guna95,logi05,logi08,ivan05,gemm13}).
\par
An alternative approach to the construction of self-duality equations proposed in Ref.~\cite{tchr80}, was to consider self-duality relations between higher-order terms of the field strength. An explicit example of instantons satisfying such self-duality relations was obtained on the eight-dimensional sphere in Ref.~\cite{gros84}. As shown in Ref.~\cite{duff91}, these instantons play a role in smoothing out the singularity of heterotic string soliton solutions by incorporating one-loop corrections. In Refs.~\cite{duff97,olsen00,mina01,pedd08,bill09,fuci09} these exotic solutions were used to construct various string and membrane solutions. In Refs.~\cite{bern03,hase14}, they were used to construct the higher dimensional quantum Hall effect.
\par
In this letter, we study the ADHM construction of the self-dual instantons in eight dimensions and we find a multi-instanton solution generalizing the solution that was found in Ref.~\cite{gros84}.

\section{Preliminaries}

In this section, we give a brief summary of Clifford algebra and octonion algebra. We list the features of the mathematical structure as far as they are of relevance to our work.
\par
We recall that the Clifford algebra $Cl_{0,7}(\mathbb{R})$ is a real associative algebra generated by the elements $\Gamma_1,\Gamma_2,\dots,\Gamma_7$ and defined by the relations
\begin{equation}\label{1}
\Gamma_i\Gamma_j+\Gamma_j\Gamma_i=-2\delta_{ij}.
\end{equation}
The element $\Omega=\Gamma_1\Gamma_2\dots \Gamma_7$ commutes with all other elements of the algebra, and its square $\Omega^2=1$. Therefore the pair $\Gamma^{\pm}=\frac{1}{2}(1\pm\Omega)$ forms a complete system of mutually orthogonal central idempotents, and hence the algebra $Cl_{0,7}(\mathbb{R})$ decomposes into the direct sum of two ideals. It can be shown that these ideals are isomorphic to the algebra $M_{8}(\mathbb{R})$ of all real matrices of size $8\times 8$. The latter, in turn, is the algebra of endomorphisms of the octonion algebra. Let us describe these relations in a little more detail.
\par
The algebra of octonions $\mathbb O$ is a real linear algebra with the canonical basis $1,e_{1},\dots,e_{7}$ such that
\begin{equation}\label{2}
e_{i}e_{j}=-\delta_{ij}+c_{ijk}e_{k},
\end{equation}
where the structure constants $c_{ijk}$ are completely antisymmetric and nonzero as $c_{123}=c_{145}=c_{167}=c_{246}=c_{275}=c_{374}=c_{365}=1$. The algebra of octonions is not associative but alternative, i.e. the associator $(x,y,z)=(xy)z-x(yz)$ is totally antisymmetric in $x,y,z$. Consequently, any two elements of $\mathbb O$ generate an associative subalgebra. The algebra of octonions satisfies the identity $((zx)y)x=z(xyx)$ which is called the right Moufang identity. The algebra $\mathbb O$ permits the involution (i.e. an anti-automorphism of period two) $x\to\bar x$ such that the elements $t(x)=x+\bar x$ and $n(x)=\bar{x}x$ are in $\mathbb R$. In the canonical basis, this involution is defined by $\bar e_{i}=-e_{i}$. It follows that the bilinear form $(x,y)=\frac12(\bar xy+\bar yx)$ is positive definite and defines an inner product on $\mathbb O$. It is easy to prove that the quadratic form $n(x)$ permits the composition $n(xy)=n(x)n(y)$. Since this form is positive definite, it follows that $\mathbb O$ is a division algebra.
\par
Denote by $R_{x}$ the operator of right multiplication on $x$ in the octonion algebra, i.e. $yR_{x}=yx$ for all $y\in\mathbb{O}$. The set of all such operators generates a subalgebra in the algebra $\text{End}\,\mathbb{O}$ of endomorphisms of the linear space $\mathbb{O}$. Using the multiplication law (\ref{2}) and antisymmetry of the associator $(e_{i},e_{j},e_{k})$, we prove the equalities
\begin{equation}\label{3}
R_{e_{i}}R_{e_{j}}+R_{e_{j}}R_{e_{i}}=-2\delta_{ij}1_8,
\end{equation}
where $1_8$ is the identity $8\times 8$ matrix. Comparing (\ref{3}) with (\ref{1}), we see that the correspondence $\Gamma_{i}\to R_{e_{i}}$ can be extended to the homomorphism $Cl_{0,7}(\mathbb R)\to\text{End}\,\mathbb O$. Since the algebra $M_{8}(\mathbb{R})$ is simple, it follows that this mapping is surjective and therefore $\text{End}\,\mathbb O\simeq M_8(\mathbb R)$. Note also that the product
\begin{equation}\label{4}
R_{e_1}R_{e_2}\dots R_{e_7}=1_8.
\end{equation}
This equality follows from simplicity of $M_{8}(\mathbb{R})$ and the fact that the element $\Omega$ lies in the center of $Cl_{0,7}(\mathbb R)$.
\par
Along with the operator of right multiplication, one can define the operator $L_x$ of left multiplication on $x$ in $\mathbb{O}$, namely $yL_{x}=xy$ for $y\in\mathbb{O}$. Suppose $\text{End}(\mathbb {O})^{(-)}$ is a Lie algebra obtained by introducing the commutator multiplication on the vector space $\text{End}(\mathbb{O})$. Its subalgebra generated by all operators right and left multiplications on $\mathbb{O}$ is called the Lie multiplication algebra of $\mathbb{O}$ and denoted by $\text{Lie}(\mathbb{O})$. It is well known (see, e.g., the review \cite{baez02}) that this algebra is generated by operators only right multiplications on $\mathbb{O}$ and is isomorphic to $so(8)\oplus\mathbb{R}1_8$. Using the antisymmetry of $(x,a,b)$, it is easy to prove the identity
\begin{equation}\label{40}
R_aR_b=R_{ab}-[R_a,L_b].
\end{equation}
Note that the identity (\ref{40}) and other relations in this paper rely on the right operator action. For example, $R_a R_b$ sends $x$ to $(xa)b$. In order to go to the left operator action, it is enough to do redefine $R_a\leftrightarrow L_a$. This follows, in particular, that the product $R_aR_b\in\text{Lie}(\mathbb{O})$ for any $a,b\in\mathbb{O}$. Note also that
\begin{equation}\label{14}
R_{\bar a}=|a|^2R_{a^{-1}},\quad R_{a^{-1}}=R_{a}^{-1},\quad R_aR_bR_a=R_{aba},
\end{equation}
where $|a|^2=\bar{a}a$. It follows from the homomorphism $Cl_{0,7}(\mathbb{R})\to \text{End}(\mathbb{O})$ that $R_{x}\in SO(8)$ if and only if $|x|=1$. Since the norm of $a/|a|$ is equal to 1 as $a\ne0$, we get
\begin{equation}\label{45}
R_x^t=R_x^{-1},\quad \det R_x=1,\quad R_{\bar{a}}=R_a^t.
\end{equation}
where $x=a/|a|$ and $0\ne a\in\mathbb{O}$. The last equality in (\ref{14}) is easily proved using the first two formulas in (\ref{14}) and (\ref{45}).
\par
Now we redenote the unit of $\mathbb{O}$ by the symbol $ e_0 $ so that $ R_ {e_0} = 1_8 $ and consider the combinations
\begin{equation}
R_{\mu\nu}=R_{e_{\mu}}\bar{R}_{e_{\nu}}-R_{e_{\nu}}\bar{R}_{e_{\mu}},
\end{equation}
where $\bar{R}_{e_{\mu}}\equiv R_{\bar{e}_{\mu}}$ and the Greek index takes values from 0 to 7. Then it easily follows from the identity (\ref{4}) that
\begin{equation}\label{5}
R_{[\mu\nu}R_{\lambda\rho]}=\frac{1}{4!}\varepsilon_{\mu\nu\lambda\rho\alpha\beta\gamma\sigma}
R_{[\alpha\beta}R_{\gamma\sigma]}.
\end{equation}
Hence, the tensor $R_{[\mu\nu}R_{\lambda\rho]}$ satisfies the self-duality equations.

\section{ADHM construction}

Let $M_{m,n}(\mathbb{R})$ be the set of all real $m\times n$ matrices, $\widetilde{K}$ a linear subspace of $\text{End}\,\mathbb{O}$, and $M$ a real $8m\times 8n$ matrix. We call $M$ the $\widetilde{K}$-matrix of size $m\times n$ if it is representable as a matrix with elements from $\widetilde{K}$, i.e. if $M\in M_{m,n}(\mathbb{R})\otimes\widetilde{K}$. In the case when $m=n$ and $M\in\mathbb{R}1_n\otimes\widetilde{K}$, where $1_n$ is the identity $n\times n$ matrix, we say that the matrix $M$ is diagonal over $\widetilde{K}$. Finally, we say that this matrix is real over $\text{End}\,\mathbb{O}$ if $M\in M_{m,n}(\mathbb{R})\otimes\mathbb{R}1_8$.
\par
Now let $K=\{R_a\mid a\in\mathbb{O}\}$. We choose two $K$-matrices $C$ and $D$ of size $(n + N)\times N$ in such a way that for any $x\in K$ the matrix
\begin{equation}\label{8}
M(x)=Cx+D
\end{equation}
satisfies the following conditions:
\par\medskip
i) the matrix $\bar{M}^{t}M$ is real over $\text{End}\,\mathbb{O}$;
\par
ii) the matrix $\bar{C}^{t}M(\bar{M}^{t}M)^{-1}\bar{M}^{t}C$ is real over $\text{End}\,\mathbb{O}$;
\par
iii) the matrix $M$ has the maximal rank $8N$.
\par\medskip\noindent
Here $\bar{M}$ signifies the transposition of elements in the blocks, $M^t$ signifies the transposition of the blocks, and $\bar{M}^t$ signifies the transposition of the real matrix. Note also that everywhere below we say that the matrix is real, implying that it is real over $\text{End}\,\mathbb{O}$.
\par
In addition to $K$, we need one more example of subspace of $\text{End}\,\mathbb{O}$. Suppose $\widetilde{K}$ is the Lie algebra generated by $K$, i.e. $\widetilde{K}$ is the Lie multiplication algebra of $\mathbb{O}$. Then it follows from (\ref{40}) that $M(x)$ is a $\widetilde{K}$-matrix of size $(n+N)\times N$. This construction will be used in Appendix A to calculate the gauge group of $N$-instanton.
\par
Now we choose a matrix $U(x)$ of size $(8n+8N)\times 8n$ over $\mathbb{R}$ such that
\begin{equation}\label{9}
\bar{U}^{t}M=0,\quad \bar{U}^{t}U=1_n
\end{equation}
and define the linear potential
\begin{equation}\label{7}
A_{\mu}(x)=\bar{U}^{t}\partial_{\mu}U.
\end{equation}
Then it follows from conditions i) and ii) that the corresponding completely antisymmetric 4-tensor $F_{[\mu\nu}F_{\lambda\rho]}$ will have the form
\begin{equation}\label{6}
F_{[\mu\nu}F_{\lambda\rho]}
=\bar{U}^{t}CR_{[\mu\nu}R_{\lambda\rho]}(\bar{M}^{t}M)^{-1}\bar{C}^{t}U\bar{U}^{t}C(\bar{M}^{t}M)^{-1}\bar{C}^{t}U.
\end{equation}
To prove this equality, it suffices to use the conditions (\ref{9}) and the reality of $\bar{C}^tC$ that follows from the conditions i). Obviously, the tensor (\ref{6}) satisfies the self-duality equations.
\par
We note that the use of the octonion algebra in this construction is not necessary. Instead of the algebra $\text{End}\,\mathbb{O}$, we can use the matrix algebra $M_{8}(\mathbb{R})$ which is isomorphic to $\text{End}\,\mathbb{O}$. In this case, it suffices to construct the homomorphism $Cl_{0,7}(\mathbb{R})\to M_{8}(\mathbb{R})$ and find the images of generators of the Clifford algebra in $M_{8}(\mathbb{R})$. This idea was made in ref.~[26], where an expression for $F_{[\mu\nu}F_{\lambda\rho]}$ coinciding with (\ref{6}) as $n=1$ was obtained.
\par
It is shown in Appendix A that the gauge group of the instanton (\ref{6}) is $Spin(8n)$. Therefore, when replacing $U$ with $UT$, where $T\in Spin(8n)$, the potential (\ref{7}) undergoes a gauge transformation, and the $N$-instanton (\ref{6}) does not change. Similarly, the $N$-instanton will be invariant under the transformations
\begin{equation}
M\to XMY,\quad U\to XU,
\end{equation}
where $X\in O(8n+8N)$ and $Y\in GL(N,\mathbb{R})\otimes\mathbb{R}1_8$. Therefore, without loss of generality, we can assume that the matrix (\ref{8}) has the form
\begin{equation}\label{10}
M(x)=\begin{pmatrix}\Lambda\\B-x1_{N}\end{pmatrix},
\end{equation}
where $\Lambda$ and $B$ are constant $K$-matrices and $x\in K$.
\par
It follows from the condition i) that the matrices $\bar{B}^tB+\bar{\Lambda}^t\Lambda$ and $\bar{B}^tx+\bar{x}B$ must be real. On the other hand, $\bar{B}^tx+\bar{x}B$ is real if and only if $B$ is symmetric. Therefore, the matrix $\bar{B}^tB+\bar{\Lambda}^t\Lambda$ is also symmetric. Generally speaking, the same $N$-instanton can be obtained from different matrices $\Lambda$ and $B$. Suppose $P$ is a special orthogonal $(8n\times8n)$-matrix, and $S\in O(N)\otimes\mathbb{R}1_8$. Then the transformation
\begin{equation}\label{44}
\Lambda\to P\Lambda S,\quad B\to S^tBS
\end{equation}
leads only to a replacement of bases and therefore leaves the $N$-instanton unchanged. Moreover, blocks of $\bar{B}^tB+\bar{\Lambda}^t\Lambda$ and $S$ are scalar $8\times8$-matrices. Therefore, by virtue of the theorem on reduction to principal axes, the matrix $\bar{B}^tB+\bar{\Lambda}^t\Lambda$ can be considered diagonal over $K$.
\par
In order to modify the condition ii), we note that
\begin{align}\label{22}\notag
&\bar{C}^tM(\bar{M}^{t}M)^{-1}\bar{M}^tC\\
&=B(\bar{M}^{t}M)^{-1}\bar{B}^t-\{B\bar{x}(\bar{M}^{t}M)^{-1}+(\bar{M}^{t}M)^{-1}x\bar{B}^t\}+(\bar{M}^{t}M)^{-1}|x|^2.
\end{align}
If the matrix $B$ is non-degenerate then the reality condition of the first term on the right hand side is equivalent to the reality of the matrix
\begin{equation}\label{23}
(\bar{B}^t)^{-1}(\bar{M}^{t}M)B^{-1}=(\bar{B}^t)^{-1}\{(\bar{B}^tB+\bar{\Lambda}^t\Lambda)-(\bar{x}B+\bar{B}^tx)+|x^2|\}B^{-1}.
\end{equation}
Since the matrix $B$ is symmetric, this condition is satisfied if and only if the matrices
\begin{equation}\label{11}
B\bar{B}\quad\text{and}\quad B(\bar{B}B+\bar{\Lambda}^t\Lambda)^{-1}\bar{B}
\end{equation}
are real.
\par
Now we find the reality condition of the second term on the right hand side of (\ref{22}). Suppose
\begin{equation}\label{20}
X=B\bar{x}(\bar{M}^{t}M)^{-1}+(\bar{M}^{t}M)^{-1}x\bar{B}\quad\text{and}\quad Y=(\bar{M}^{t}M)X(\bar{M}^{t}M).
\end{equation}
\par\noindent
Obviously, $X$ is real iff $Y$ is real. Using (\ref{10}), we represent $Y$ in the form
\begin{align}
&x\bar{B}(\bar{B}B+\bar{\Lambda}^t\Lambda)+(\bar{B}B+\bar{\Lambda}^t\Lambda)B\bar{x}\label{27}\\
&-x\bar{B}(\bar{x}B+\bar{B}x)-(\bar{x}B+\bar{B}x)B\bar{x}\label{25}\\
&+(x\bar{B}+B\bar{x})|x|^2.\label{21}
\end{align}
We need the following two simple statements:
\par\medskip
a) the matrix $R_aR_{\bar{b}}+R_bR_{\bar{a}}$ is real for any $a,b\in\mathbb{O}$;
\par\medskip
b) the equality $R_aR_{\bar{b}}+R_bR_{\bar{a}}=R_{\bar{b}}R_a+R_{\bar{a}}R_b$ is true for any $a,b\in\mathbb{O}$.
\par\medskip\noindent
In order to prove the statements, it is enough to use the identity $(ya)b+(yb)a=y(ab+ba)$, and notice that $\bar{a}=2a_0-a$ and $\bar{b}=2b_0-b$, where $a_0,b_0\in\mathbb{R}$. As a result, we get $ab+\bar{b}\bar{a}=ba+\bar{a}\bar{b}$ and
\begin{equation}
R_aR_{\bar{b}}+R_bR_{\bar{a}}=R_{4a_0b_0-ab-\bar{b}\bar{a}},
\end{equation}
that proves the statements.
\par
It follows from a) that the matrix (\ref{21}) is real. In order to prove the reality of (\ref{25}), we consider the matrix
\begin{equation}\label{26}
x\bar{B}(\bar{x}B+\bar{B}x)+B\bar{x}(\bar{x}B+\bar{B}x).
\end{equation}
It is real by the statement a). Summing (\ref{25}) and (\ref{26}), and using the reality of $\bar{x}B+\bar{B}x$, we get
\begin{equation}\label{30}
A(x)=\{B(\bar{x}B+\bar{B}x)-(\bar{x}B+\bar{B}x)B\}\bar{x}.
\end{equation}
Obviously, $A(x)$ is real iff (\ref{25}) is real. As noted above, the matrix $B$ is symmetric and the matrix $S=\bar{B}B+\bar{\Lambda}^t\Lambda$ is real and diagonal over $K$. Therefore $B\bar{B}=(\bar{B}B)^t=\bar{B}B$. Using this equality and the reality of $B\bar{B}$, we transform (\ref{30}) to the form
\begin{equation}\label{28}
A(x)x\bar{B}=|x|^2\{(B\bar{x}+x\bar{B})-(\bar{x}B+\bar{B}x)\}B\bar{B}.
\end{equation}
It follows from b) that the expression in curly brackets is zero. Therefore, if the matrix $\bar{B}$ is non-degenerate, then $A(x)=0$. This proves the reality of the matrix (\ref{25}).
\par
Similar considerations apply to the matrix (\ref{27}). It is enough to replace $\bar{x}B+\bar{B}x$ with $\bar{B}B+\bar{\Lambda}^t\Lambda$. As a result, we obtain an analogue of (\ref{30})
\begin{equation}
\tilde{A}(x)=\{(\bar{B}B+\bar{\Lambda}^t\Lambda)B-B(\bar{B}B+\bar{\Lambda}^t\Lambda)\}\bar{x},
\end{equation}
where $\tilde{A}(x)$ is again a real matrix for any $x\in K$. Substituting $x=x_0$, we prove the reality of the expression in curly brackets. But then $\tilde{A}(\vec{x})$, where $\vec{x}=\frac{1}{2}(x-\bar{x})$, cannot be real. Hence, the matrix (\ref{27}) is real iff
\begin{equation}\label{29}
(\bar{B}B+\bar{\Lambda}^t\Lambda)B=B(\bar{B}B+\bar{\Lambda}^t\Lambda).
\end{equation}
Note that the reality of the second matrix in (\ref{11}) is a consequence of (\ref{29}) and the reality of $B\bar{B}$ or $\bar{\Lambda}^t\Lambda$.
\par
As for the condition iii), it is satisfied if and only if the equations $B\xi=x\xi$ and $\Lambda \xi=0$ have a unique solution $\xi=0$ for any $x\in K$. Thus, we obtain the following equivalent reformulation of conditions i), ii), and iii) (under the condition that the matrix $B$ is non-degenerate):
\par\medskip
i)$'$ the matrix $B$ is symmetric, and the matrix $\bar{B}B+\bar{\Lambda}^t\Lambda$ is real and diagonal over $K$;
\par
ii)$'$ the matrices $B$ and $\bar{B}B+\bar{\Lambda}^t\Lambda$ are mutually commutative and the matrix $\bar{\Lambda}^t\Lambda$ is real;
\par
iii)$'$ the equations $B\xi=x\xi$ and $\Lambda \xi=0$ has a unique solution $\xi=0$ for any $x\in K$.
\par\medskip
Unlike the conditions i) and ii), all matrices in i)$'$ and ii)$'$ are constant, which greatly simplifies the research of the moduli space. It is difficult to find the module parameters directly from condition ii), because these parameters must satisfy the constraints for an arbitrary $x$. That is why the multi-instanton solutions in Ref.~\cite{naka16}, where the condition ii)$'$ is absent, were obtained only for a limited situation when each instanton is well separated.

\section{Multi-instanton solutions}

Let us consider in more detail the case $n=1$, when the potential (\ref{7}) takes values in the Lie algebra $so(8)$. To construct an $N$-instanton satisfying the conditions (\ref{9}), we will search for the matrix $U=U(x)$ in the following form
\begin{equation}\label{43}
U=k\begin{pmatrix}-1_8\\V\end{pmatrix},
\end{equation}
where the column vector $V=(v_1,\dots,v_N)^{t}$, $v_i\in K$ and the real $k>0$. Substituting this expression into the formula (\ref{7}) and using the conditions (\ref{9}), we get the expressions
\begin{equation}\label{15}
A_{\mu}=\frac{1}{2}\frac{\bar{V}^t\partial_{\mu}V-\partial_{\mu}\bar{V}^tV}{1+\bar{V}^tV},
\end{equation}
where $\bar{V}^t=\Lambda(B-x1_8)^{-1}$ and $1+\bar{V}^tV=k^{-2}$. Note that potentials (\ref{15}) and (\ref{7}) as $n=1$ are gauge equivalent.
\par
Now back to the consideration of the conditions i)$'$, ii)$'$, and iii)$'$. Suppose $\Lambda=(\lambda_1,\dots,\lambda_N)$ and $B=(b_{ij})$, where $b_{ij}=b_{ji}$. Then the diagonal elements of $\bar{B}^tB+\bar{\Lambda}^t\Lambda$ have the form
\begin{equation}
k_j=\sum_{i=1}^N|b_{ij}|^2+|\lambda_{j}|^2
\end{equation}
and, therefore, are real. Therefore, condition i)$'$ is reduced to the $N(N-1)/2$ relations
\begin{equation}\label{31}
\sum_{i=1}^N\bar{b}_{ij}b_{ik}+\bar{\lambda}_{j}\lambda_{k}=0,\qquad j<k,
\end{equation}
which expresses the vanishing of off-diagonal elements of $\bar{B}^tB+\bar{\Lambda}^t\Lambda$.
\par
In order to rewrite condition ii)$'$, we consider the transformation $\Lambda\to T\Lambda$, where $T\in Spin(7)$. Since the group $ Spin (7) $ acts transitively on the set of elements of the norm 1 in $K$, we can assume that $\lambda_1$ is real and positive. On the other hand, the matrix $\bar{\Lambda}^t\Lambda$ is real and elements of her first row have the form $\lambda_1\lambda_i$. Therefore, all elements of $\Lambda$ must be real. Hence, condition ii)$'$ is reduced to the relations
\begin{equation}\label{32}
(k_i-k_j)b_{ij}=0,\qquad\lambda_i\in\mathbb{R}1_8,\quad\lambda_1>0.
\end{equation}
(Here and everywhere below we write $\lambda_1>0$, implying that $\lambda_1=k1_8$ with $k>0$.) Thus, if $b_{ij}=0$, then the additional condition do not arise. If $b_{ij}\ne0$, then  we have the additional condition
\begin{equation}\label{33}
\sum_{k=1}^N|b_{ik}|^2+\lambda_{i}^2=\sum_{k=1}^N|b_{jk}|^2+\lambda_{j}^2.
\end{equation}
\par
Finally, the condition iii)$'$ is equivalent to the following requirement. For any $x\in K$, the system of equations
\begin{equation}
\sum_{j=1}^Nb_{ij}\xi_j=x\xi_i,\qquad \sum_{i=1}^N\lambda_{i}\xi_i=0
\end{equation}
has only the zero solution $\xi_i=0$.
\par
We turn to the study of special cases. Obviously, the conditions i)$'$, ii)$'$, and iii)$'$ are automatically satisfied in the 1-instanton case. Let $N=2$. If $b_{12}=0$, then it is obvious that $\lambda_1>0$ and $\lambda_2=0$. While $b_{11}$ and $b_{22}$ are independent variables. Suppose $b_{12}\ne 0$. Then it follows from (\ref{31}) and (\ref{32}) that
\begin{equation}\label{13}
\bar{b}_{11}=-\bar{b}_{12}b_{22}b_{12}^{-1}-\lambda_{1}\lambda_{2}b_{12}^{-1},
\end{equation}
where $\lambda_2$ is real and $\lambda_1>0$. This equation has a solution in the space $K$. Indeed, using the identities (\ref{14}), we prove that the first term on the right-hand side of (\ref{13}) belongs to $K$. The second term is in $K$ because $\lambda_1$ and $\lambda_2$ are real. Hence, $\bar{b}_{11}\in K$, as it should be. For $N=2$, the condition (\ref{33}) takes the form
\begin{equation}\label{34}
|b_{11}|^2+\lambda_{1}^2=|b_{22}|^2+\lambda_{2}^2.
\end{equation}
In this case, $\lambda_1>0$, $b_{22}$, and $b_{12}\ne0$ can be selected as independent elements.
\par
In order to satisfy condition iii)$'$, we exclude element $\xi_1$ from system 1. We obtain
\begin{align}
(x\lambda_1^{-1}\lambda_2-b_{11}\lambda_1^{-1}\lambda_2+b_{12})\xi_2&=0,\\
(x+b_{12}\lambda_1^{-1}\lambda_2-b_{22})\xi_2&=0.
\end{align}
Hence, the condition iii)$'$ is satisfied if and only if for any $x$, at least one of the coefficients at $\xi_2$ is nonzero. It is easy to see, that this is equivalent to the condition
\begin{equation}
b_{11}-b_{12}\lambda_1\lambda_2^{-1}\ne b_{22}-b_{12}\lambda_1\lambda_2
\end{equation}
as $\lambda_2\ne0$, and the the condition $x\ne b_{22}$ as $b_{12}=\lambda_2=0$. Thus, the ansatz (\ref{15}) defines a 2-instanton in the following cases:
\par
1) the elements $\lambda_1>0$, $b_{11}$ and $b_{22}$ are independent variables, and $b_{12}=\lambda_2=0$;
\par
2) the elements $\lambda_1>0$, $b_{22}$ and $b_{12}\ne0$ are independent variables, and $b_{11}$ and $\lambda_2$ are defined by the formulas (\ref{13}) and (\ref{34}) respectively.
\par
Obviously, in both cases there are 17 free real parameters.
\par
The case $N=3$ is investigated in a similar way. Suppose $B$ is a non-degenerate symmetric matrix of size $3\times3 $ over $K$. Then condition i)$'$ reduces to solving the system of equations
\begin{align}\label{18}
\bar{b}_{11}b_{12}+\bar{b}_{12}b_{22}+\bar{b}_{13}b_{23}+\lambda_{1}\lambda_{2}&=0,\notag\\
\bar{b}_{11}b_{13}+\bar{b}_{12}b_{23}+\bar{b}_{13}b_{33}+\lambda_{1}\lambda_{3}&=0,\\
\bar{b}_{12}b_{13}+\bar{b}_{22}b_{23}+\bar{b}_{23}b_{33}+\lambda_{2}\lambda_{3}&=0.\notag
\end{align}
Consider the following four possibilities.
\par
1) Let $b_{12}=b_{13}=b_{23}=0$. Setting $\lambda_1>0$, we have $\lambda_2=\lambda_3=0$. The conditions ii)$'$ is satisfied automatically. The conditions iii)$'$ is satisfied if $b_{11}+b_{22}+b_{33}\ne0$. Thus, we have 25 free real parameters in total.
\par
2) Let $b_{12}\ne0$ and $b_{13}=b_{23}=0$. Setting $\lambda_1>0$, we have $\lambda_3=0$. So we reduce the system under consideration to one equation equivalent to (\ref{13}). Arguing as above, we prove that the elements $b_{12}\ne0$, $b_{22}$, $b_{33}$ and $\lambda_1>0$ are independent variables, and $b_{11}$ and $\lambda_2$ are defined by the formulas (\ref{13}) and (\ref{34}) respectively. Thus, we again have a total of 25 free real parameters.
\par
3) Let $b_{12}\ne0$, $b_{13}\ne0$ and $b_{23}=0$. We put choose the elements $b_{12}\ne0$, $b_{22}$ and $\lambda_1>0 $ as independent variables. Through them, $\bar{b}_{11} $, $b_{13}$ and $b_{33}$ are easily expressed
\begin{align}
\bar{b}_{11}&=-\bar{b}_{12}b_{22}b_{12}^{-1}+\lambda_{1}\lambda_{2}b_{12}^{-1},\label{35}\\
b_{33}&=-\bar{b}_{13}^{-1}\bar{b}_{11}b_{13}+\lambda_{1}\lambda_{3}\bar{b}_{13}^{-1},\label{38}\\
b_{13}&=-\lambda_{2}\lambda_{3}\bar{b}_{12}^{-1}.\label{37}
\end{align}
It follows from (\ref{14}) that solutions of the equations must be belongs to $K$. To satisfy ii)$'$, we impose the conditions
\begin{align}
|b_{11}|^2+\lambda_{1}^2&=|b_{22}|^2+\lambda_{2}^2,\label{36}\\
|b_{22}|^2+\lambda_{2}^2&=|b_{33}|^2+\lambda_{3}^2.\label{39}
\end{align}
The variables $b_{11}$ and $\lambda_2$, we find by solving the system (\ref{35}) and (\ref{36}). Substituting (\ref{37}) in (\ref{38}), we get $b_{33}$. Then (\ref{39}) and (\ref{37}) give $\lambda_3$ and $b_{13}$. As a result, we obtain a solution with 17 free real parameters.
\par
4) Let $b_{12}\ne0$, $b_{13}\ne0$, and $b_{23}\ne0$. Then the first equation in (\ref{18}) has the solution $b_{11}\in K$ only if $b_{23}=kb_{13}$, where $k\in\mathbb{R}$. We rewrite the other two equations in the form of a system with respect to the unknown $b_{33}$ and $\lambda_3$. This system will have a solution in $K$ only if $b_{13}=\lambda_1$ and $\lambda_2\ne k\lambda_1$. The conditions (\ref{36}) and (\ref{39}) fix the values of $\lambda_2$ and $k$. Hence, the independent variables are $b_{12}\ne0$, $b_{22}$ and $\lambda_1>0$. As a result, we again obtain a solution with 17 free real parameters.
\par
We see that the number of free parameters does not increase with an increase in the number of nonzero off-diagonal elements of $B$. This is due to the need to have solutions in the space $K$. This is a very strong condition. It can be shown that a similar situation takes place in the case of arbitrary $N$. Indeed, let $B_0$ be a non-degenerate symmetric $N\times N$ matrix over $K$ in which all off-diagonal elements are zero. Then the independent variables of the system (\ref{31}) are $b_{11},\dots,b_{NN}$ and $\lambda_1$. Now let $B_1$ be a non-degenerate matrix that differs from $B_0$ only in the element $b_{jk}\ne0$ with $j<k$. It follows from $(j,k)$-th equation in (\ref{31}) that $\bar{b}_{jj}$ and $b_{kk}$ are linearly dependent. Hence $B_0$ and $B_1$ contain the same number of independent elements. Using induction on the number of nonzero off-diagonal matrix elements, we consider the transition from $B_i$ to $B_{i+1}$, where again $B_{i+1}$ differs from $B_i$ only in the off-diagonal element $b_{jk}\ne0$. Suppose $b_{jk}$ and $b_{jl}$, where $j<l$, are independent elements. Then $(k,l)$-th equation in (\ref{31}) contain the arbitrary term $\bar{b}_{jk}b_{jl}$. In order for this equation to have a solution in the space $K$, an additional condition is necessary.
\par
Thus, to exclude the additional conditions occurrence, it is necessary to choose off-diagonal elements with non-matching indexes. In this case, as is easy to see, the number of independent elements of $B$ cannot be more than $N$. Now note that together with $b_{jk}\ne0$, an additional condition (\ref{33}) arises that binds the variables $\lambda_j$ and $\lambda_k$. Therefore, the number of independent elements of $\Lambda$ should also not increase. Thus, we have proved that the $N$-instanton in eight dimensions satisfying the conditions i), ii), and iii) cannot have more than $8N+1$ free parameters.
\par
A solution containing exactly $8N+1$ free parameters is easy to construct explicitly. Suppose that the $N\times N$ matrix $B$ is non-degenerate and diagonal over $K$, and the matrix $\Lambda=(\lambda_1,\dots,\lambda_N)$, where all $\lambda_i>0$. Then the conditions i$'$) is satisfied for a suitable orthogonal transformation $B\to S^{-1}BS$ and $\Lambda\to\Lambda S$, the conditions ii$'$) is satisfied when choosing $\sum_i\lambda_is_{ij}=0$ for $j=2\dots N$, and the conditions iii$'$) is true as $x\ne b_i\equiv b_{ii}$. Substituting the values of $\Lambda$ and $B$ in (\ref{15}), we find the potential
\begin{equation}\label{19}
A_{\mu}=\frac{1}{2}R_{\nu\mu}\partial_{\nu}\ln\left(1+\sum^{N}_{i=1}\frac{\lambda_i^2}{|b_i-x|^2}\right),
\end{equation}
which is an eight-dimensional analogue of the 't Hooft instanton in dimension four. In particular, for $N=1$, the obtained instanton is gauge equivalent (outside the point $x=b_1$) to the 1-instanton that was found in Ref.~\cite{gros84}. The singularities in the gauge field at $x=b_i$ are not physical but are artifacts of our choice $v_0=-1_8$ in (\ref{43}).
\par
Following Ref.~\cite{gros84} and regarding the field strength $F=\{F_{\mu\nu}\}$ as a curvature, we define the topological charge \begin{equation}
Q=k\int\text{tr}(F\wedge F\wedge F\wedge F),
\end{equation}
for a suitable normalization constant k, as the fourth Chern number. It was shown in Ref.~\cite{naka16}, that the topological charge so defined is quantized in the well-separated limit, i.e. if $|b_i-b_j|^2\gg\lambda_i\lambda_j$ for all $i\ne j$. Therefore such instanton can be regarded as a superposition of $N$ instantons and hence for it the topological charge coincides with $N$. On the other hand, if $Q$ is the topological charge of the 't Hooft type instanton with the well-separated limit, we can argue that $Q$ must be the topological charge for any 't Hooft type solution since the latter can be obtained from the solution with the well-separated limit by continuous deformation of $B$ and $\Lambda$. Thus, the topological charge of the instanton (\ref{19}) is $N$.

\section{Conclusion and discussions}

In this article, we have studied the ADHM construction of self-dual instantons in eight dimensions, which was proposed in Ref.~\cite{naka16}. Using the well-known connection between the Clifford algebra $Cl_{0,7}(\mathbb{R})$ and the octonion algebra $\mathbb{O}$, we found new restrictions on matrices in instead of those used in the cited work. This made it possible to calculate the dimension of the space of solutions to this construction in the case $n=1$ and found a new $N$-instanton solution of the 't~Hooft type. In addition, we have shown that the gauge group of the theory is $Spin(8n)$.
\par
Since the moduli space  $W_n(N)$ here is the quotient space of the space of all pairs $(\Lambda,B)$ satisfying conditions i)$'$-iii)$'$ with respect to the equivalence relation given by (\ref{44}), it contains an open everywhere dense subset $W'_n(N)$, which is a smooth manifold. We have shown that $\dim W'_1(N)=8N+1$. Unfortunately, the construction of charts of $W'_1(N)$ is complicated by the need to take into account factorization with respect to relations (\ref{44}). Therefore, we have built charts (or independent parameters of $N$-instantons) only for manifolds $W'_1(2)$ and $W'_1(3)$. We stress that our method does not work for $N>3$, since equations (\ref{31}) become nonlinear and, of course, all these calculations say practically nothing about the topology of $W_1(N)$.
\par
Unlike the approximate solution of the 't~Hooft types constructed in Ref.~\cite{naka16}, the solution (\ref{19}) is exact. It containing $8N+1$ free parameters and its topological charge is $N$. The singularities in (\ref{19}) are not physical but are artifacts of our choice of the singular gauge. For $N=1$, the instanton (\ref{19}) is gauge equivalent to the 1-instanton that was found in Ref.~\cite{gros84}. Note also that the projection of $\mathbb{O}$ onto its subalgebra $\mathbb{H}$ transforms (\ref{19}) into the 't~Hooft instanton in dimension four.

\appendix
\section{Gauge group of the instanton}

Let $G$ be a group Lie, $A_G$ its Lie algebra, and the potential $A_{\mu}(x)=\bar{U}^{t}\partial_{\mu}U$ takes values in $A_G$. The gauge transformation of $A_{\mu}(x)$ is induced by the transformation $U\to UT$, where $\bar{T}^{t}=T^{-1}$ and $T\in G$. On the other hand, it follows from (\ref{40}) that the product $\bar{U}^{t}\partial_{\mu}U$ takes values in the Lie multiplication algebra $\widetilde{K}$. Since the latter is isomorphic to $so(8)\oplus\mathbb{R}1_8$ it follows that $G\simeq Spin(8)$ as $n=1$.
\par
In the case $n>1$, the situation is complicated by the fact that the commutator of two $\widetilde{K}$-matrices, generally speaking, is not a $\widetilde{K}$-matrix. Therefore, it is only obvious that $G$ is a subgroup of $Spin(8n)$. However, this statement can be strengthened. For this purpose, we introduce several new designations. Let us denote by the symbols $e_{\{ps\}}$ and $e_{[ps]}$  $\widetilde{K}$-matrices with the elements
\begin{equation}
(e_{\{ps\}})_{p's'}=\delta_{pp'}\delta_{ss'}+\delta_{ps'}\delta_{sp'},\quad
(e_{[ps]})_{p's'}=\delta_{pp'}\delta_{ss'}-\delta_{ps'}\delta_{sp'},
\end{equation}
and let $R_{[ij\dots k]}=R_{e_i}R_{e_j}\dots R_{e_k}$, where $i<j<\dots<k$.
\par
Further, let $\widetilde{V}$ be a linear space generated by the $\widetilde{K}$-matrices
\begin{equation}\label{41}
R_ie_{\{ps\}},\quad R_{[ij]}e_{\{ps\}},
\end{equation}
where $i,j=1,2\dots7$ and $p,s=1,2\dots n$. Obviously, these elements are a basis of the space. In addition, any $v\in\widetilde{V}$ satisfies the condition $\bar{v}^t=-v$. Hence $\widetilde{V}$ is a subspace of $so(8n)$. Suppose $A_G$ is a Lie algebra generated by $\widetilde{V}$. Then $A_G$ is a subalgebra of $so(8n)$. On the other hand, the commutators
\begin{equation}\label{42}
[R_ie_{\{ps\}},R_ie_{\{pp\}}]=2e_{[ps]},\quad [R_ie_{\{pp\}},R_{[jk]}e_{\{ps\}}]=2R_{[ijk]}e_{[ps]}.
\end{equation}
The elements (\ref{41}) and (\ref{42}) are linearly independent and belong to $A_G$, and their number is
\begin{equation}
14{n(n+1)}+18{n(n-1)}=\dim{so(8n)}.
\end{equation}
Hence it follows that $A_G$ coincides with $so(8n)$ and therefore the gauge group $G\simeq Spin(8n)$ or $SO(8n)$.
\par
Note that the ADHM construction does not impose the specialty condition on the gauge group in general. Therefore, the choice of $Spin(8n)$ as the gauge group seems more preferable in comparison with $SO(8n)$.

\end{document}